\documentclass[12pt,aps,prd,nofootinbib,preprintnumbers]{revtex4}
\pdfoutput=1

\usepackage[utf8]{inputenc}
\usepackage{graphicx}
\usepackage{amsfonts}
\usepackage{latexsym}
\usepackage{amsmath,amssymb}
\usepackage{float}
\def\a2{\ddot{a}}

\begin{document}

\title{Ricci cosmology }
\author{ Rudolf Baier} 

\affiliation{Faculty of Physics, University of 
	Bielefeld, D-33501 Bielefeld, Germany\\ baier@physik.uni-bielefeld.de}

\author{Sayantani Lahiri}

\affiliation{Universit\"at Bremen -
	Center of Applied Space Technology and Microgravity (ZARM)
	Am Fallturm, 28359 Bremen, Germany\\
sayantani.lahiri@zarm.uni-bremen.de}

\author{Paul Romatschke}

\affiliation{Department of Physics and\\ Center for Theory of Quantum Matter - University of Colorado, Boulder, CO 80309, USA\\
paul.romatschke@colorado.edu}

\preprint{BI-TP 2019/04}

\date{\today}

\begin{abstract}
  \vspace{2cm}
  We revisit spatially flat FLRW cosmology in light of recent advances in standard model relativistic fluid dynamics. Modern fluid dynamics requires the presence of curvature-matter terms in the energy-momentum tensor for consistency. These terms are linear in the Ricci scalar and tensor, such that the corresponding cosmological model is referred to as ``Ricci cosmology''. No cosmological constant is included, there are no inflaton fields, bulk viscosity is assumed to be zero and we only employ standard Einstein gravity. Analytic solutions to Ricci cosmology are discussed, and we find that it is possible to support an early-time inflationary universe using only well-known ingredients from the Standard Model of physics and geometric properties of space-time.
\vspace{8cm}	
\end{abstract}

\maketitle


\section{Introduction}

The $\Lambda$CDM model is the accepted concordance model for modern day precision cosmology. Unfortunately, neither the cold dark matter (CDM) component nor the dark energy $\Lambda$ are understood in terms of the modern Standard Model of physics. While dark matter also features prominently in astrophysics such as in explaining galaxy rotation curves, dark energy is more mysterious, and apparently only is needed to support an expanding universe.

An elegant possibility to support expansion using only Standard Model ingredients is that of so-called bulk viscous cosmology, cf. Refs.~\cite{Murphy:1973zz,Padmanabhan:1987dg,Zimdahl:1996ka,Normann:2016zby,Normann:2016jns}. In bulk viscous cosmology, the perfect fluid energy-momentum tensor is supplemented by a viscous term that couples to the local expansion rate, generating an effective pressure that can be lower than the pressure of a perfect fluid. Assuming extreme cases of the bulk viscosity coefficient, the effective pressure can be made negative, thereby driving an expanding universe. While bulk viscous cosmology supports a de Sitter universe using only well-known ingredients, two significant challenges have so far prevented bulk viscous cosmology to replace the $\Lambda$CDM model. First, the extreme values of bulk viscosity necessary to support an expanding universe are hard to generate using only ingredients from the Standard Model of physics, and second, bulk viscous cosmology seems hard to reconcile with precision cosmology data \cite{Li:2009mf,Gagnon:2011id,Velten:2011bg,Velten:2012uv}. This provides the motivation to consider other potential alternatives to dark energy that support a de Sitter universe with only Standard Model ingredients.

In this work, we consider corrections to the perfect fluid energy-momentum tensor at \textit{second} order in gradients. Generalizing the ideas of M\"uller, Israel and Stewart \cite{Muller:1967zza,Israel:1979wp}, these terms naturally arise in modern derivations of fluid dynamics as an effective field theory of low-momentum modes \cite{Baier:2007ix,Bhattacharyya:2008jc}, and indeed are \textit{required} in order to ensure consistency of two-point correlation functions (cf. Ref.~\cite{Romatschke:2017ejr} for a modern textbook on the subject.) Since some of these second order gradient terms are those of second order gradients of the metric (curvature terms), the terms constitute generic corrections to the perfect fluid energy-momentum tensor even in local thermodynamic equilibrium. In this sense, they differ in physical origin from the more familiar bulk viscous correction terms. However, like in bulk viscous cosmology, the curvature terms can reduce the effective pressure of the fluid (cf. section 2.2.3 in Ref.~\cite{Romatschke:2017ejr}), and therefore could offer a potential alternative mechanism for inflation. This provides the motivation for the present work.

\section{Setup}

Let us consider the four-dimensional Einstein-Hilbert action including matter given by
\begin{eqnarray}
S[g_{\mu \nu}]&=& \frac{1}{2}\int d^4 x \sqrt{-g} R +\underbrace{\int d^4 x \sqrt{-g} {\cal L}_{M}}_{S_{matter}}\,,  \label{Action}
\end{eqnarray}
where $\sqrt{-g}$ is the determinant of the metric $g_{\mu \nu}$ and we are using units where Newton's constant is set to $G=\frac{1}{8\pi}$. For a homogeneous, isotropic and spatially flat spacetime we consider the FLRW metric
\cite{dInverno:1992gxs,Poisson:2009pwt}
\begin{equation}
  \label{metric}
ds^2= -dt^2+a^2(t)\delta_{ij}dx^{i}dx^{j}\,,
\end{equation} 
where $a(t)$ is the scale factor characterizing the cosmological evolution of the universe \cite{Baumann:2009ds}. Conventional cosmology assumes the matter contribution to be given by the so-called ``perfect fluid'' form \cite{LL}
\begin{equation}
  \label{eq:pf}
T_{\mu \nu}^{\rm perfect\, fluid}=\rho u_{\mu}u_{\nu}+p\left (u_\mu u_\nu+g_{\mu \nu}\right)\,,
\end{equation}
where $\rho$ is the energy density, $p$ is the fluid pressure and $u_\mu$ is the fluid four velocity. In the context of cosmology, it is often stated that (\ref{eq:pf}) is the most general expression consistent with the FLRW symmetries for a fluid in local thermodynamic equilibrium. However, while true for flat space-times, this statement is not correct when space-time is curved. 

Driven largely by the need to provide theoretical input to the experimental high-energy physics program at colliders, the theory of relativistic fluid dynamics has seen considerable advances in the past decades \cite{Romatschke:2017ejr}. Using the tools from effective field theory, fluid dynamics has been set up in terms of a gradient expansion of the fundamental degrees of freedom $\rho, u_\mu, g_{\mu\nu}$, which allows a rigorous derivation of all possible terms in the energy momentum tensor to a given order in gradients, cf. Refs.~\cite{Baier:2007ix,Bhattacharyya:2008jc,Grozdanov:2015kqa} for conformal fluids.

From the perspective of relativistic fluid dynamics, at the second order in the gradient expansion, it has been shown that for non-conformal fluids in equilibrium the energy-momentum tensor takes the form \cite{Romatschke:2017ejr,Romatschke:2009kr}
\begin{equation}
  \label{vis-fluid}
  T_{\mu \nu}=\rho u_{\mu}u_{\nu}+p_{\rm eff}\left (u_\mu u_\nu+g_{\mu \nu}\right)\,,\quad
  p_{\rm eff}=p+\xi_5 R+\xi_6 u^\alpha u^\beta R_{\alpha\beta}\,,
\end{equation}
where $R_{\alpha \beta}$ is the Ricci tensor, $R$ is the Ricci scalar and $p_{\rm eff}$ is the effective pressure. Here $\xi_{5}, \xi_{6}$ are ``second-order'' transport coefficients (see appendix \ref{appa}). Note that in this work we focus on effects arising from curvature terms only, and hence other (in particular non-equilibrium) terms have been neglected in Eq.~(\ref{vis-fluid})\footnote{Also recall that for isotropic backgrounds, shear viscous terms do not contribute to the dynamical evolution of the universe.}. Also note that the resulting dynamics of cosmology should be distinguished from the Ricci dark energy model described in Ref.~\cite{Zimdahl:2014yza}.

The effective pressure in Eq.~(\ref{vis-fluid}) only includes terms up to second order in derivatives, leading to the question if (\ref{vis-fluid}) is only valid for small corrections $|p-p_{\rm eff}|\ll p$. Indeed, since higher order derivative terms are allowed by the symmetries, and one might wonder how reliable the truncation of Eq.~(\ref{vis-fluid}) at second-order will be for real-world systems. To answer this question, one would need a result for the energy-momentum tensor for ``all-orders fluid dynamics'', which for a quantum-field theory in curved space-time is presently unknown. However, results for all-orders fluid dynamics do exist in more restricted settings, driven by recent developments of hydrodynamic attractors and resurgence \cite{Heller:2015dha,Romatschke:2017vte,Aniceto:2018uik,Denicol:2018pak}. In brief, in cases where the full, all-orders result for fluid dynamics is known, it has been found that low-order truncations (e.g. at first or second-order) provide a good quantitative approximation to the full, non-perturbative result even when gradients are large.  In this work, we assume that a similar attractor exists also for fluid dynamics in curved space-time, and that Eq.~(\ref{vis-fluid}) provides a reasonable approximation to this attractor even when $|p-p_{\rm eff}|>p$.

Using the FLRW metric (\ref{metric}) in the comoving frame, we have
\begin{eqnarray}
  u^{\alpha}u^{\beta}R_{\alpha \beta}=R_{00}&=&-\displaystyle \frac{\ddot{a}}{a}=-3(\dot{H}+H^2)\,,\nonumber\\
  R&=&6 \displaystyle\left(\frac{\ddot{a}}{a}+\frac{\dot{a}^2}{a^2}\right)=6\displaystyle(\dot{H}+2H^2)\,,
\end{eqnarray}
in terms of the Hubble function $H(t)= \displaystyle\frac{\dot{a}}{a}$. Variation of the action (\ref{Action}) with respect to $g_{\mu \nu}$ gives rise to Einstein equation,
\begin{equation}
G_{\mu \nu}=R_{\mu \nu}-\frac{1}{2}R g_{\mu \nu} =T_{\mu \nu}~, \label{EE}
\end{equation} 
and substituting the FLRW metric together with (\ref{vis-fluid}) in the (\ref{EE}) gives rise to the Friedmann equation and ``acceleration equation'' as 
\begin{equation}
  \label{friedmann}
  \left(\displaystyle\frac{\dot{a}}{a}\right)^2 = H^2=\frac{\rho}{3}\,,\quad
  \dot{H}=-\displaystyle \frac{1}{2}\left(\rho+p_{\rm eff}\right)\,.
  \end{equation}
It is worthwhile mentioning here that (\ref{friedmann}) is linear in the Ricci terms compared to the cosmological model proposed by
Starobinsky \cite{Starobinsky:1980te}.

Dimensional analysis reveals that for $p_{\rm eff}$ given in Eq.~(\ref{vis-fluid}), the transport coefficients $\xi_{5,6}\sim \cal O(\sqrt{\rho})$, such that the Friedmann equation allows us to write
\begin{equation}
  \xi_{5,6}=\hat\xi_{5,6} H\,,
\end{equation}
with dimensionless $\hat\xi_{5,6}$. Furthermore, we write the equation of state as $p=w \rho$ such that Eq.~(\ref{friedmann}) can be written as
\begin{equation}
  \label{master}
  \dot{H}(1+c_1 H)+c_{0}H^2+c_2H^3=0\,,
\end{equation}
with $c_0 = \frac{3 (1+w)}{2}$, $c_1=3\hat{\xi}_5-\frac{3}{2}\hat{\xi}_6$, $c_2 =6\hat{\xi}_5-\frac{3}{2}\hat{\xi}_6$. Note that the effective pressure in this case becomes
\begin{equation}
  \label{eq:peff}
  p_{\rm eff}=\frac{w+\left(\frac{2 c_2}{3}-c_1\right)\sqrt{\rho/3}}{1+c_1 \sqrt{\rho/3}}\rho\,,
  \end{equation}
which for non-vanishing $c_{1,2}$ is non-linear in the density $\rho$, cf. Ref.~\cite{Zimdahl:2000zm}. We stress that Eq.~(\ref{eq:peff}) is not a model, but instead a consequence from standard  quantum field theory and fluid dynamics.

\section{Simple Solutions in Ricci Cosmology}

In the following, we will be interested in a qualitative analysis of Ricci cosmology. For this reason, we will study solutions to (\ref{master}) assuming a simple form of the equation of state of the matter and the transport coefficients $\xi_{5,6}$. While a more realistic treatment will be necessary in order to study the viability of the model in describing precision cosmology observations, we will leave this task for future study. Specifically, we make the assumptions
\begin{equation}
  w={\rm const.}\geq 0\,,\quad \hat\xi_{5,6}={\rm const.}\,,
\end{equation}
such that the coefficients $c_0,c_1,c_2$ in (\ref{master}) are pure numbers and $c_0>\frac{3}{2}$. Note that we only consider equations of state for ordinary matter for which $w\geq 0$, whereas for dark energy generically $w<0$. Also, in the following we will only consider the case $H(t)>0$, even though a similar analysis of (\ref{master}) could be applied to gravitational collapse.

\subsection{Standard Cosmology without a Cosmological Constant}
\label{sec:stan}

	As reference, let us first consider the case with vanishing transport coefficients i.e. $ {\xi}_5 = {\xi}_6 = 0$ implying $c_1 = c_2 = 0$.  Then Eq.~(\ref{master}) reduces to
	\begin{eqnarray}
	\dot{H}+c_0 H^2 =0\,,
	\end{eqnarray}
which	for $w \neq -1$ and initial conditions $H(t=0)=1$, $a(t=0)=1$  leads to a decelerating universe, 
	\begin{equation}
	  H(t)=\frac{1}{(c_0 t +1)}\,,\quad
          a(t)=(1+c_0 t)^{1/c_0}\,.\label{ref}
	\end{equation}
        For this solution $\frac{\ddot{a}}{a} = (1 - c_{0}) (1 + c_{0} t)^{-2}<0$ since $c_0>\frac{3}{2}$, so in this case the universe is decelerating. Note that for $w=0$ ($c_0=\frac{3}{2}$), this corresponds to the solution of the Einstein-de Sitter model \cite{ORaifeartaigh:2015qkt}, and for the case $w=-1$ ($c_0=0$) this corresponds to the de Sitter model where $H={\rm const}.$

        \subsection{Simple Ricci Cosmology Solutions}

        \begin{figure}[t]
          \centering
\includegraphics[width=0.9\linewidth]{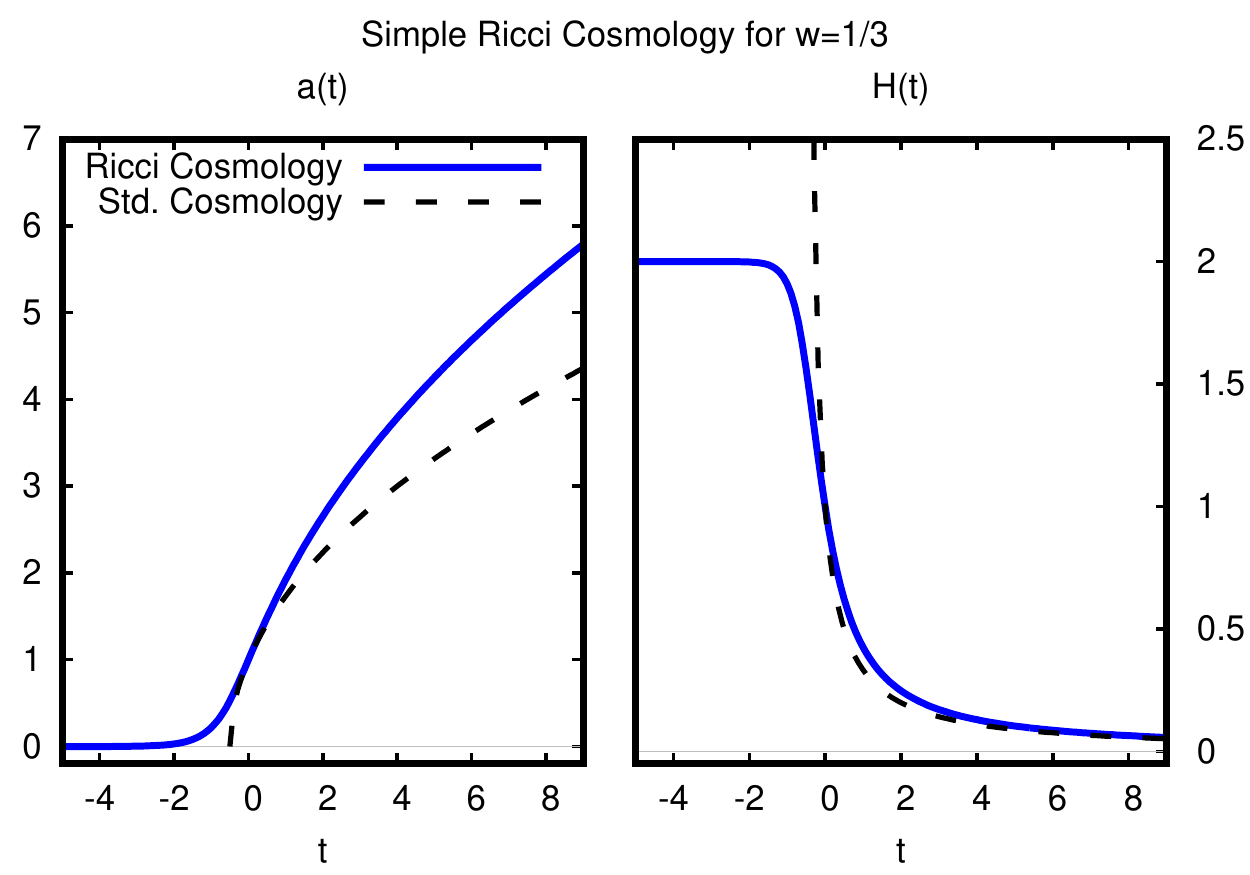}
\caption{Ricci cosmology with $c_1=0$, $c_2=-\frac{3 (1+w)}{4}$ for radiation $w=\frac{1}{3}$ compared to standard cosmology ($c_1=c_2=0$) for radiation, Eq.~(\ref{ref}). Left panel shows the evolution of the scale factor, while the right panel shows the Hubble function. Thin line is a guide to the eye. Note that Ricci cosmology includes an inflationary phase at early times, avoiding the singularity in standard cosmology. \label{fig1}}
\end{figure}

Let us consider the case when one of the three coefficients is vanishing. We take $c_1=0$ giving rise to $\hat{\xi}_5= \displaystyle \frac{1}{2}\hat{\xi}_6$ such that we obtain,
\begin{equation}
\frac{\dot{H}}{H^2} + c_0 + c_2 H = 0\,.
\label{basiciii}
\end{equation}
Note that Eq.~(\ref{basiciii}) matches a similar equation that appeared in the context of bulk viscous cosmology investigated by Murphy \cite{Murphy:1973zz} using $\zeta = \alpha \rho$ if we set $c_2 = -\frac{9 \alpha}{2}$. Let us thus study (\ref{basiciii}) for the case of negative $c_2$, e.g. $c_2 = 3 \hat{\xi}_5 = - \vert c_2 \vert$. The solution of Eq.(\ref{basiciii}) in terms of the scale parameter $a(t)$ with $a(t = 0) = 1$ reads
\begin{equation}
  \frac{c_0^2 t}{\vert c_2 \vert}  + 1 = c_0 \ln a + a^{c_0}\,,\quad
  H=\frac{c_0}{|c_2|(1+a^{c_0})}\,.
  \label{simsol}
\end{equation}
For small scale factor $a(t)\ll 1$, this case gives the de Sitter metric,
\begin{equation}
a(t) \approx \exp{\left(\frac{c_0 t}{\vert c_2 \vert}\right)}\,,
\end{equation}
while on the other hand for $a(t)\gg 1$ we have 
\begin{equation}
a(t) \approx [{c_0^2 t /{\vert c_2 \vert} + 1}]^{1/c_0}\,.
\end{equation}
This is nothing else but the Einstein-de Sitter model discussed above in (\ref{ref}). The numerical solution for $a(t),H(t)$ from Eq.~(\ref{basiciii}) for $w=1/3$ and initial condition $H(t=0)=1, a(t=0)=1$ (implying $c_2=-c_0/2$) is shown in Fig.~\ref{fig1}. One finds that Ricci cosmology contains an inflationary early-time phase that smoothly goes over to a standard radiation-dominated universe. The early-time singularity of standard cosmology with pure radiation is avoided.

It is straightforward to consider similar simple solutions with $c_1=0$ for Ricci cosmology for other forms of matter, such as dust ($w=0$). We find that these behave qualitatively the same as long as $c_2$ is negative.

The combination  $\rho+3 p_{\rm eff}$ is negative for early times in Ricci cosmology, which is the underlying physical reason for the inflationary phase, cf. Eq.~(\ref{eq:dota}).

\subsection{De Sitter Solutions for Ricci Cosmology}

Let us now consider all three coefficients to be non-vanishing. In this case Eq.~(\ref{master}) allows an exact de Sitter solution with $\dot H=0$ for $c_2<0$, namely $H(t)=H_0=-\frac{c_0}{c_2}$, independent from $c_1$.
 We therefore have for $c_2<0$,
\begin{equation}
a(t) \propto e^{H_0 t}\,,
\end{equation}


The Hubble flow parameters $\epsilon_{H}$ and $\eta_{H}$ are given by \cite{Brevik:2017msy}
\begin{eqnarray}
  \label{flows}
\epsilon_{H}& =& -\frac{\dot{H}}{H^2} = \frac{c_0 + c_2 H}{1 + c_1 H} ~, \label{ep-H} \nonumber \\
\eta_{H} &=&  \left(\frac{\ddot{H}}{\dot{H}H}-\frac{2\dot{H}}{H^2} \right) = \frac{\dot{\epsilon}_{H}}{H \epsilon_{H}}= \frac{\dot{H}(c_2-c_1 c_0)}{H(c_0+c_2 H)(1+c_1 H)}~.
\end{eqnarray}
The spectral index $n_s$ and the tensor to scalar ratio $r$ in terms of $\epsilon_{H}$ and $\eta_{H}$ can be expressed as  \cite{Brevik:2017msy, Liddle:2000cg}
\begin{eqnarray}
1-n_s &=& 2 (\epsilon_{H} + \eta_{H})~, \nonumber \\
\Rightarrow \qquad n_s &=& 1-2 \left(\frac{c_0 + c_2 H}{1 + c_1 H}+\frac{\dot{H}(c_2-c_1 c_0)}{H(c_0+c_2 H)(1+c_1 H)}\right)~, \\
 \text{and} \qquad r&=& 16 \epsilon_{H} \nonumber \\
\hspace{2cm} &=& 16  \left(\frac{c_0 + c_2 H}{1 + c_1 H} \right)~.
\end{eqnarray} 
Such definitions of $n_s$ and $r$ are valid so long as the Hubble flow parameters $\epsilon_{H}$ and $\eta_{H}$ are small during inflation. It may be seen that from following expressions,
\begin{equation}
  \label{eq:dota}
\frac{\ddot{a}(t)}{a(t)} = H^2 ( 1 - \epsilon_H)=-\frac{1}{6}\left(\rho+3 p_{\rm eff}\right),\qquad \eta_{H}=
\frac{\dot{\epsilon}_{H}}{H \epsilon_{H}}~,
\end{equation}
that the universe ceases to accelerate when $\epsilon_{H}=1$ and $\dot{a}=$ constant. This corresponds to the time when the combination $\rho+3 p_{\rm eff}=0$.

In general, the de Sitter solution corresponds to $\epsilon_{H}=0$ and $r=0$ implying exact scale invariance of curvature perturbations.

\subsection{More General Solutions for Ricci Cosmology}

 \begin{figure}[t]
          \centering
\includegraphics[width=0.9\linewidth]{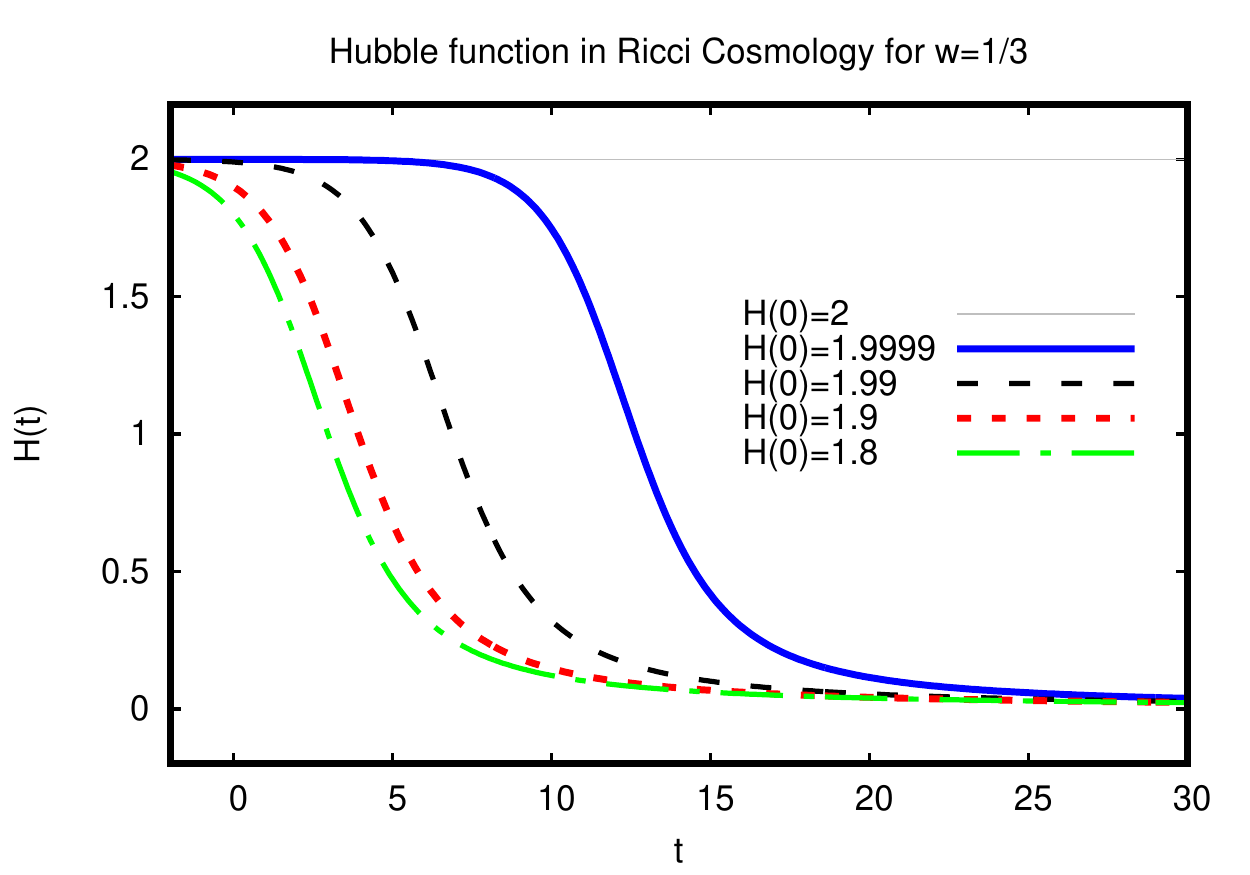}
\caption{Ricci cosmology with $c_1=2$, $c_2=-\frac{3 (1+w)}{4}$ for radiation $w=\frac{1}{3}$ for different initial conditions $H(0)$. The case of $H(0)=-\frac{c_0}{c_2}=2$ corresponds to the exact de Sitter solution (thin line). \label{fig2}}
\end{figure}

We may consider solutions to Eq.~(\ref{master}) which are close to, but not exactly equal to the de Sitter solution. The full solution to Eq.~(\ref{master}) for positive Hubble parameter $H>0$  is given by
\begin{equation}
 c_0 t + \frac{1}{H(0)} = \frac{1}{H(t)} + \left(c_1 -\frac{c_2}{c_0}\right)
\ln\left[{\frac{c_0 + c_2 H(t)}{H(t)(c_0/H(0) + c_2)}}\right]\,,
\label{solvi}
\end{equation}
where $H(0)$ is an integration constant corresponding to the value of the Hubble function at $t=0$. We note that for $c_1 = 0$, the solution to Eq.~(\ref{solvi}) coincides with Eq.~(\ref{simsol}). Furthermore for the special case $c_1 = \frac{c_2}{c_0}$, note that Eq.~(\ref{solvi}) corresponds to standard cosmology discussed in section \ref{sec:stan}, whereas for $H(0)=-\frac{c_0}{c_2}$ and $c_2<0$ we obtain the exact de Sitter solutions discussed above.

Eq.~(\ref{solvi}) may be studied for values of $H(0)\neq -\frac{c_0}{c_2}$, shown in Fig.~\ref{fig2}. Since for these solutions $\dot H\neq 0$, Eq.~(\ref{flows}) implies a non-zero value for $\epsilon_{H}$, which would be necessary in order to match the observational constraints from baryon acoustic oscillations (BAO) \cite{Li:2018iwg}.

\begin{figure}[t]
          \centering
\includegraphics[width=0.9\linewidth]{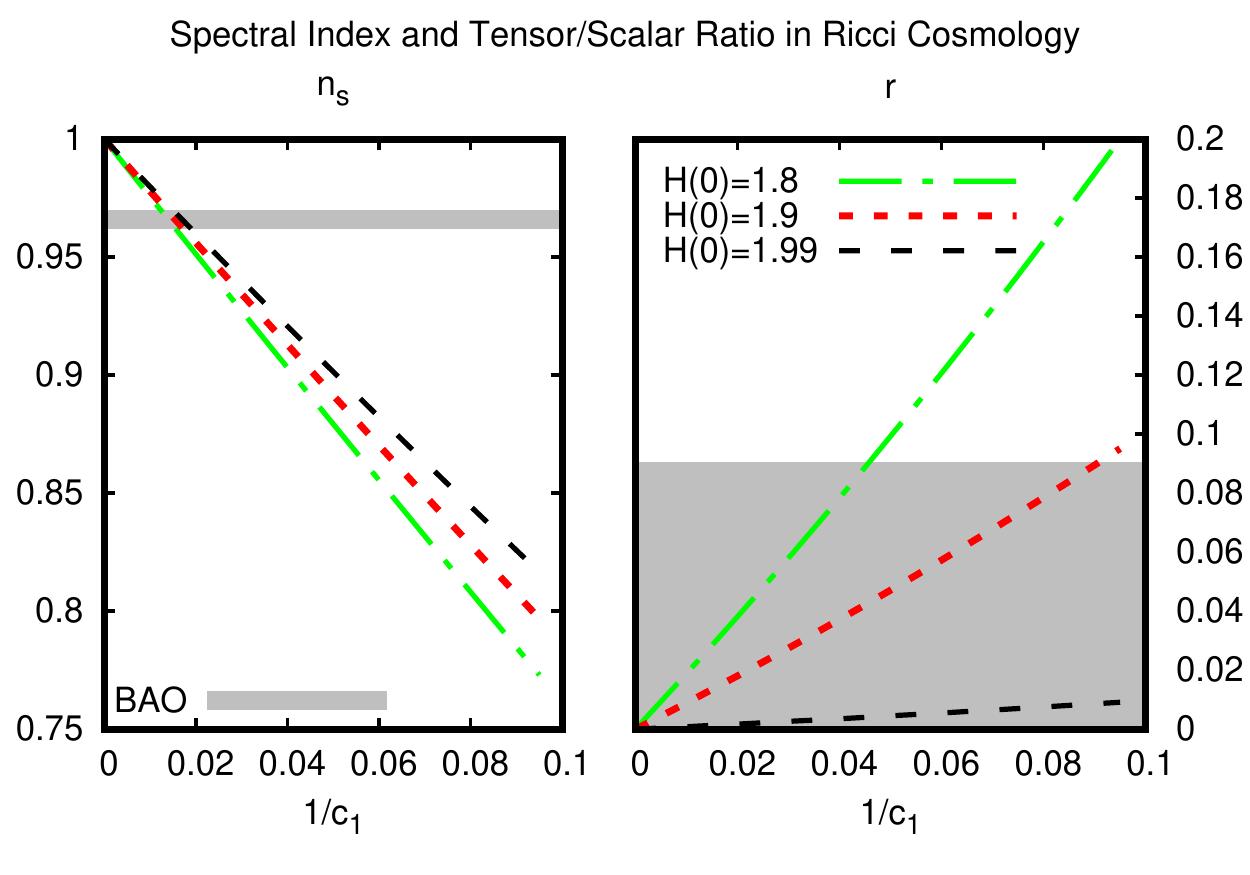}
\caption{Spectral index and Tensor/Scalar ratio in Ricci cosmology at $t=1$ as a function of $c_1^{-1}$ with $c_2=-\frac{3 (1+w)}{4}$ for radiation $w=\frac{1}{3}$ for different initial conditions $H(0)$. For reference, the current constraints from BAO observations \cite{Li:2018iwg} are indicated as filled bands. \label{fig3}}
\end{figure}

For these 'almost de Sitter solutions', the values of the spectral index $n_s$ and the tensor-to-scalar ratio $r$, calculated at $t=1$ for different initial conditions and values of $c_1$,  can be seen in Fig.~\ref{fig3}. From this figure, it can be seen that Ricci cosmology is able to comply with the BAO constraints rather generically.

\section{Summary and Conclusions}

In summary, including the curvature-matter coupling contributions to the cosmic fluid energy-momentum tensor leads to additional terms in the Einstein equations for cosmology. We stress that these terms are a consequence from quantum field theory and fluid dynamics, and are firmly grounded in the Standard Model of Physics. Whenever matter is present, these additional terms modify the equations of motion for cosmology, though interesting effects are limited to the early universe when matter densities are large.

In this work, we found that under certain conditions (when the coefficient $c_2$ introduced above is negative), the resulting solutions for Ricci cosmology can give rise to an inflationary phase for the early universe, without the presence of dark energy or bulk viscosity. The coefficient $c_2$ requires the calculation of second-order transport coefficients for general quantum field theories, and only incomplete results exist to date. Further work will be required in order to make definite statements about the sign of $c_2$.

We considered solutions with simplified assumptions (purely radiative matter, simple transport coefficients) to study the qualitative aspects of Ricci cosmology. In order to investigate if Ricci cosmology could potentially become a candidate for real-world cosmology, these assumptions will have to be lifted, which will require numerical solutions of the resulting equations. Also, a more detailed comparison of Ricci cosmology with observational data would be required. We leave these studies for future work.

\acknowledgments

We would like to thank D.~Schwarz for helpful comments on the manuscript. This work was supported, in part, by the US Department of Energy, DOE award No. DE-SC0017905 and by the German Research Foundation DFG, grant/40400957 and grant/40401089.

\appendix

\section{Transport coefficients}
\label{appa}

While generic results for the second order transport coefficients $\xi_5$ and $\xi_6$ do not exist, results for specific models are available (cf. Tab.2.2 in Ref.~\cite{Romatschke:2017ejr}). For instance, in a gauge/gravity model $\xi_5,\xi_6$ for matter at finite temperature $T$ have been obtained as 
\begin{equation}
\xi_5 = \frac{\kappa}{3} ( 1 - 3 c_s^2), ~~ \xi_6 = \frac{\xi_5}{c_s^2}\,,
\label{c1}
\end{equation}
where $\kappa \propto s/T$, $s$ is the entropy density and $c_s$ is the (constant) speed of sound  \cite{Kanitscheider:2009as}.

Writing $s \propto T^3$ gives $\kappa \propto T^2  \propto \sqrt{\rho}$.
For $ s= \frac{(\rho + p)}{T} = \frac{(1 + w)\rho}{T}$, one observes
$\kappa > 0$ for $w > -1$.
As long as $c_s^2 < 1/3$ the coefficients $\xi_{5,6}$ are positive quantities
for
\begin{equation}
-1 < w < 1/3.
\label{e2}
\end{equation}
However, the sign of these coefficients is dependent on the field content. For instance, for purely scalar fields \cite{Moore:2012tc,Kovtun:2018dvd}, the coefficient $\kappa$ is negative for all coupling values, e.g. 
\begin{equation}
  \kappa \propto - N T^2\,,  
\end{equation}
in the vector $O(N)$ model \cite{Romatschke:2019gck}. Therefore, either sign for $\xi_5,\xi_6$ is possible.

\bibliography{ricci}

\end{document}